\documentclass[fleqn,usenatbib]{mnras}
\usepackage{mathptmx}
\usepackage[T1]{fontenc}
\usepackage{ulem}
\usepackage{color}    
\DeclareRobustCommand{\VAN}[3]{#2}
\let\VANthebibliography\thebibliography
\def\thebibliography{\DeclareRobustCommand{\VAN}[3]{##3}\VANthebibliography}

\usepackage{graphicx}	
\usepackage{amsmath}	
\usepackage{amssymb}	
\usepackage[switch]{lineno}

\usepackage{xcolor}     

\title [Accretion geometry]{Accretion geometry in neutron star low-mass X-ray binaries during the hard spectral state}
\author[E. Meyer-Hofmeister, Y. L. Wang and B. F. Liu]{
Emmi Meyer-Hofmeister$^{1}$, 
Yilong Wang$^{2,3}$ 
and 
B. F. Liu$^{2,3}$\thanks{bfliu@nao.cas.cn}\\
$^{1}$Max-Planck-Institut f\"ur Astrophysik, Karl-Schwarzschild-Str.~1, D-85741 Garching, Germany\\
$^{2}$National Astronomical Observatories, Chinese Academy of Sciences, 20A Datun Road, Beijing 100101, China\\
$^{3}$School of Astronomy and Space Science, University of Chinese Academy of Sciences, 19A Yuquan Road, Beijing 100049, China
}

\date{Accepted XXX. Received YYY; in original form ZZZ}

\pubyear{2025}
\begin{document}
\label{firstpage}
\pagerange{\pageref{firstpage}--\pageref{lastpage}}
\maketitle

\begin{abstract}
We investigate the accretion geometry in neutron star low-mass X-ray
binaries (LMXBs) in the hard spectral state. It is commonly accepted
that, for low mass transfer rates, an advection-dominated accretion flow
(ADAF) is present in the inner region. But the observed
relativistically broadened emission lines in the reflection spectra
clearly indicate the existence of discs near the innermost stable
circular orbit $(R_{\rm{ISCO}})$. We investigate the interaction
between the coronal flow and the disc in neutron star LMXBs, and find
that gas condensation from the dominant, coronal accretion flow to an
inner disc is enhanced as compared to that in black hole LMXBs as a
consequence of irradiation of the corona by the neutron star
surface. Computations show that for low mass transfer rates ($\sim 0.005-0.02$
  Eddington rate) a persistent weak disc can coexist with a coronal flow in the innermost region, where a pure ADAF would have been expected. The inner disc extends outwards from $R_{\rm{ISCO}}$ to $\sim 10 R_{\rm{ISCO}}$ for Eddington ratios ($L/L_{\rm{Edd}}$) as low as $\sim 0.002$, covers a larger region for higher Eddington ratios, and eventually connects to the outer disc at $L/L_{\rm{Edd}} \sim 0.02$, thereby transiting to a soft state. We demonstrate that the observationally inferred region of the broad iron lines in the hard-state sources generally lies within the extension of the inner discs predicted by the condensation model. Disappearance of the broad iron lines is predicted at very low luminosities, either caused by very low accretion rates or disc truncation by strong magnetic fields.
\end{abstract} 



\section{Introduction}
In low-mass X-ray binaries (LMXBs) matter flows from a Roche-lobe
filling low-mass star via an accretion disc to the compact primary
star, a black hole or a neutron star. LMXBs can be transient systems
with changes between outburst and quiescent time intervals or  
in a persistent state \citep[for reviews, see][]{Remillard2006,Bahramian2023}. The mass  accretion 
rate determines this
behaviour and the accretion flow geometry. Generally it is accepted that
for a mass accretion rate $\dot{m}$ (scaled in
Eddington accretion rate) higher than about 0.02  a disc reaches inward to the last stable
orbit around the compact object. The radiation from the disc then
causes a soft spectrum. For lower mass accretion
rates a truncated disc
is expected together with an advection-dominated accretion flow (ADAF)
in the inner region \citep[][for review see
  \citealp{yuna2014}]{nayi1994}. The X-rays
from the innermost region cause a hard spectrum. In binaries with a
magnetized neutron star the accretion flow towards the compact object
can be halted by the magnetosphere depending on the magnetic field
strength \citep{gho1978, rapp2004, gilf2014}.

A large amount of observations for transient black hole and neutron
star LMXBs provide information about the truncation of discs, in
particular about changes between hard and soft spectral state during outbursts \citep{esin1997}. Depending on the
change of  mass accretion rate the inner edge of the disc lies at a different
distance to the centre. It is expected that the disc recedes from the
compact star if the mass accretion rate is low
\citep{tom2009,done2007}. Low level accretion in neutron star and
black hole X-ray binaries was studied by \citet{wij2015}. 

In a
sandwich-like structure with a corona above and below the disc the interaction of disc and corona is an
important feature. It can lead to 
 evaporation of matter from the underneath disc to the corona \citep[e.g.][] {mmh1994, lmmh1995, mlmh2000, Rozanska2000, liu2002} 
or condensation of matter from the corona  to the  disc  
\citep[e.g.][]{lmmh2006, mlmh2007, ltmmh2007}.  The  first process, known as evaporation, is possible in a wide range of distances from the centre,
where  thermal conduction plays a key role. The
second one occurs only near the compact object, where the external Compton scatterings become important in cooling the corona. Both features depend on the
mass flow rate, which determines the critical distance for changing from evaporation to condensation \citep[for a review, see][]{lbfq2022}. 

The disc evaporation model \citep {mmh1994, lmmh1995} originally was developed  for cataclysmic variables,
but is also valid for the formation of
the corona in  black hole LMXBs \citep{mlmh2000,liu2002} and  neutron star
LMXBs \citep{lmmh2005, mhlm2005}.
Evaporation also occurs around supermassive
black holes in active galactic nuclei \citep[AGNs; e.g.][]{liu1999,lmh2001,liu2009,taamlyq2012, mhlq2017, qlbf2018}. The evaporation  can transfer the
dominant accretion flow from a thin disc to a hot ADAF/corona for $\dot{m}$ lower than 0.02. It thus provides a  physical mechanism for
the transition of spectral states in LMXBs \citep[e.g.][]{mlmh2000}.
The condensation model \citep{lmmh2006, mlmh2007, ltmmh2007} is an
interesting supplementation to the diverse geometry of
the accretion flows. It allows to understand the observation of inner
discs in the canonical hard state. This important evidence for the
existence of thin discs around the innermost stable circular orbit (ISCO) was
found in observations with {\it XMM-Newton, Suzaku}, {\it INTEGRAL}, {\it Chandra} and {\it Swift} \citep{reis2010}.

 The interaction of disc and corona in the whole region from the outer edge
to the centre was studied in detail and was found to result in
evaporation in the outer region and condensation  in the
innermost region at intermediate accretion rates (\citealp{lmmh2006,
taam2008,taamlyq2012,qlbf2018}, for a review see \citealp{
lbfq2022}).  This leads to the formation of a truncated disc
 at large radii, an ADAF in the middle,  and a corona  coexisting with a condensation-fed  disc
at small radii (\citealp[e.g. Fig. 1 of][]{lmmh2006}, \citealp[Fig. 2 of][]{ mayer2007}). 
The radial extensions of these three regions  depend on the accretion rate.
The solutions are valid for accretion onto stellar-mass black holes and also super massive black holes in AGNs at low/intermediate accretion rates, and have been confirmed by a simplified, vertically-averaged approximation of the disc corona interaction model \citep[][]{chona2022}.

For the geometry of the accretion flows in disc and corona important
information can be deduced from the observed reflection spectra which result from external illumination of the
accretion disc by X-rays. The most prominent emission line in the
reflection spectrum is the iron $K_\alpha$ emission line
\citep[e.g.][]{ross2005}. The relativistically broadened Fe lines
are indicative of reflection from the innermost disc and allow to determine
the inner radius of the accretion disc,
and give information about other properties such as the
inclination. In a large number of studies, reflection spectra of
black hole and neutron star LMXBs in soft or hard
spectral states were discussed, for example, by \citet{cack2010}. For many
neutron star LMXBs with a magnetic field the reflection spectra from the inner disc
radii were taken to get an upper limit
of the strength of the magnetic field \citep{ibra2009, cack2009}.

For the reflection studies {\it NuSTAR} (Nuclear
Spectroscopic Telescope Array; \citealp{harr2013}) is an exceptional
tool due to its large energy bandpass from 3 to 79 keV, as well as its
high effective collecting area free from instrumental effects such as
pile-up \citep{lumb2019}. Together
with {\it NICER} (Neutron Star Interior Composition
Explorer; \citealp{gen2012}) with 
the bandpass 0.2-12 keV and high spectral sensitivity, 
the two telescopes are ideal
for revealing the presence of reflected emission while pinning down
the continuum \citep{lujg2021}.

The field of reflection modeling is discussed in a recent review by
\citet{lud2024}. Inner disc radius as a function of luminosity
 was documented for several neutron star LMXBs 
in the investigation of \citet{mou2023}. For some
binaries in hard spectral state inner disc radii were found, which is an
accretion geometry differing from
the general picture of the existence of only an ADAF
in the inner region during this spectral state. As pointed out by
\citet{lumc2017} there appears no  clear correlation trend between
mass accretion rate and the location of the inner disc radius. 

For black hole LMXBs the possible existence of weak inner discs  was
recently studied theoretically by \citet{wang2024} taking into account significant
modifications to previous work \citep[e.g.][]{qlbf2018}. Since broadened Fe lines were also
found in the reflection spectra of neutron star LMXBs during the hard
spectral state we now investigate the
re-condensation in these sources. The additional radiation from the
neutron star surface contributes to the irradiation of the accretion flow
and influences the establishment of the equilibrium between
corona and disc.

In this paper we will first give a short description of the
condensation model as it was applied to black hole LMXBs. Then
we elucidate the effect of the irradiation of the coronal flow by
the radiation from the neutron star surface. Based on the 
condensation model modified by the neutron star irradiation 
we computed the accretion flow in disc and corona. The
dependence of the results on mainly the mass flow rate and Eddington ratio 
is evaluated. By comparison with observations we show that the disc caused by the condensation of  hot gas 
can well explain the relativistically-broadened emission lines.

\section{Description of the condensation model}

The condensation is caused by the interaction of the coronal flow and
the disc flow underneath, close to the compact object. The model takes
into account the energy coupling between  the two accretion flows
by electron thermal conduction, the external Compton cooling of
the corona by the disc soft photons, and the reprocessing and
reflection of coronal irradiation. This interaction was studied for
black hole LMXBs in several investigations
\citep[e.g.][]{lmmh2006,ltmmh2007,mlmh2007,taamlyq2012,wang2024,wang2025}. 

We give a short description of the physics of the interaction between
the accretion flows in corona and disc underneath. The corona can be described by a self-similar solution  with the same form as an ADAF
\citep{nayi1995}. These solutions depend on the advection fraction of
the viscous heating, which is assumed to be about 
1 for an ADAF. But in the case with a disc underneath, the situation for the ADAF/corona is different
\citep{wang2024}. In the corona the inverse Compton scattering off the soft photons
from the disc leads to more sufficient cooling and hence
a small value of the advection fraction of
the viscous heating, which is self-consistently calculated.

In the corona the ions are directly heated by viscosity. For electrons
the heating is assumed via Coulomb collisions. The
cooling is much more complicated as it involves multiple radiative
processes in the corona, as well as the thermal conduction from the
hot corona to the cooler transition layer.
The radiative cooling includes \citep{qlbf2018} the bremsstrahlung radiation,  
the synchrotron radiation, the corresponding self-Compton and  
the external Comptonization. The total soft photon flux for the external 
Comptonization in the corona consists of the local disc
emission by viscous process and reprocessing of the coronal radiation,
and the central disc (if exists) emission.
For the irradiation flux a lamp-post model is used
which simplifies the corona as a point source above the compact object. For the detailed description we refer to the work of
\citet{wang2024}.

\section{Condensation in neutron star LMXBs}

 As compared to the case of black hole accretion, in neutron star X-ray binaries  the radiation from the neutron star
surface provides additional irradiation of the corona.
Thus, the total soft photon flux, $F_{\rm soft}(R)$, for Comptonization in the corona at
any distance  $R$,
is composed of the fluxes contributed by the local disc accretion, $F_{\rm loc}(R)$, coronal irradiation reprocessed in the disc, $F_{\rm irr}(R)$,  central disc emission, $F_{\rm cen}(R)$, and the emission from the neutron star surface, $F_{\rm sur}(R)$, 
\begin{equation}
    \begin{aligned}
    &F_{\rm loc}(R) = \frac{3GM\dot{M}_{\rm disc} (R)}{8\pi R^3}\left(1 - \sqrt{\frac{3R_{\rm S}}{R}}\right), \\
    & F_{\rm irr} (R)= \frac{(1-a)L_{\rm cor}}{8\pi}\frac{H_{\rm s}}{(R^2 + H^2_{\rm s})^{3/2}},\\
    &F_{\rm cen}(R)= 0.29\frac{L_{\rm disc}}{8\pi R^2} e^{-\tau_{\rm es}(R)},\\
    & F_{\rm sur}(R)=0.29\frac{L_*} {4\pi R^2} e^{-\tau_{\rm es}(R)},  
    \end{aligned}
    \label{eq:softflux}
\end{equation}
where $G$ is the gravitational constant, $M$ the mass of neutron star,
$\dot{M}_{\rm disc}(R)$  the local accretion rate via disc, 
 $R_{\rm S}=2GM/c^2$ the Schwarzschild radius, $H_{\rm s}$ the equivalent  height of   the corona emission
  taken in the lamp-post model,
${\tau_{\rm es}}$ the optical depth of electron scattering,
and $a= 0.15$  the albedo of the coronal irradiation. $L_{\rm disc}$,
$L_{\rm cor}$, and $L_*$ are the luminosities from disc, corona and
neutron star surface, respectively. The formulae
for $F_{\rm loc}(R)$, $F_{\rm irr} (R)$ and $F_{\rm cen}(R)$ are the same as
for condensation in black hole binaries \citep{wang2024}, where  $0.29$ in $F_{\rm cen}(R)$ is the height-averaged factor and $e^{-\tau_{\rm es}(R)}$
is the attenuation factor due to electron scattering \citep[][]{wang2024}. 
$F_{\rm cen}(R)$ is 
 dominantly contributed by the innermost region of the disc and should not be repeatedly included in the soft photons when calculating the innermost coronal flow. 

 The extra Compton cooling of the corona by the soft photons from
  the neutron star enhances the condensation of the corona, leading to
  the formation of a larger and stronger inner disc at a given accretion rate than in black hole binaries.
This effect depends on the luminosity emitted from the surface of the
neutron star, $L_*$. For accretion via a standard disc, half
  of the gravitational energy of the gas is released in the disc before reaching the surface of the neutron star $R_*$, i.e. $\frac{GM\dot M}{2R_*}\approx 0.1 \dot M c^2$.
The
  remaining half, stored as kinetic and internal energy and expected
  to be dissipated in the boundary layer of the neutron star surface
   is included with $F_{\rm sur}(R)$. Therefore, 
it is reasonable to parameterize the emission power of $L_*$ from the neutron star surface as
\begin{equation}\label{e:lambda}
L_*=\lambda (0.1 \dot M c^2),    
\end{equation} 
where $\lambda$ denotes the fraction of  kinetic and internal energy
carried to the ISCO  by  the accreted gas to be radiated at the surface of the neutron star. In the case of radiation inefficient accretion, the energy carried by the gas is usually  larger than $0.1 \dot M c^2$,  equivalent with a  somewhat smaller  $\lambda$  for a given $L_*$. 
 
Constraints to the emission fraction of the energy carried by the
accretion flow have been investigated for a pure ADAF  in low-luminosity neutron star LMXBs by comparing with  observations 
  \citep{qlbf2020,qiao2020}, which  indicates  a value
  for $\lambda$   in the range $0.01\lesssim\lambda\lesssim
    0.5$. Considering  that the two-phase accretion flows lose more
    energy to radiation in the accretion process than that of the
    ADAF, as well as potentially anisotropic radiation of the neutron
    star, photons  which can be intercepted by the corona are even less. Thus, we choose  $\lambda=0.01,0.05,0.1$ for calculations in this work. 

 As will be shown in the next section, the extra Compton
  cooling by radiation of the neutron star could lead to the
  formation of an inner disc at a luminosity as low as  $L/L_{\rm Edd}
  \sim 10^{-3}$. However, this effect becomes  less important as the accretion rate increases and hence the soft photons from the disc dominate.

\section {Properties of the accretion flows in neutron star LMXBs} 

For the numerical computations of re-condensation the values
1.4 $M_{\odot}$ and 12.5 km are taken for the mass and the radius of
the neutron star \citep{qlbf2018}. 
For the magnetic parameter $\beta$
(with magnetic pressure $p_m=B^2/8\pi=(1-\beta)p$, $p$ = magnetic
pressure + gas pressure) we assume a value of 0.95 \citep[][]{wang2024}. 
For the viscosity parameter value  $\alpha=0.3$ is taken, which is observationally constrained  \citep[see][]{king2007}.

\begin{figure}
\centering
\includegraphics [width=8.0cm]{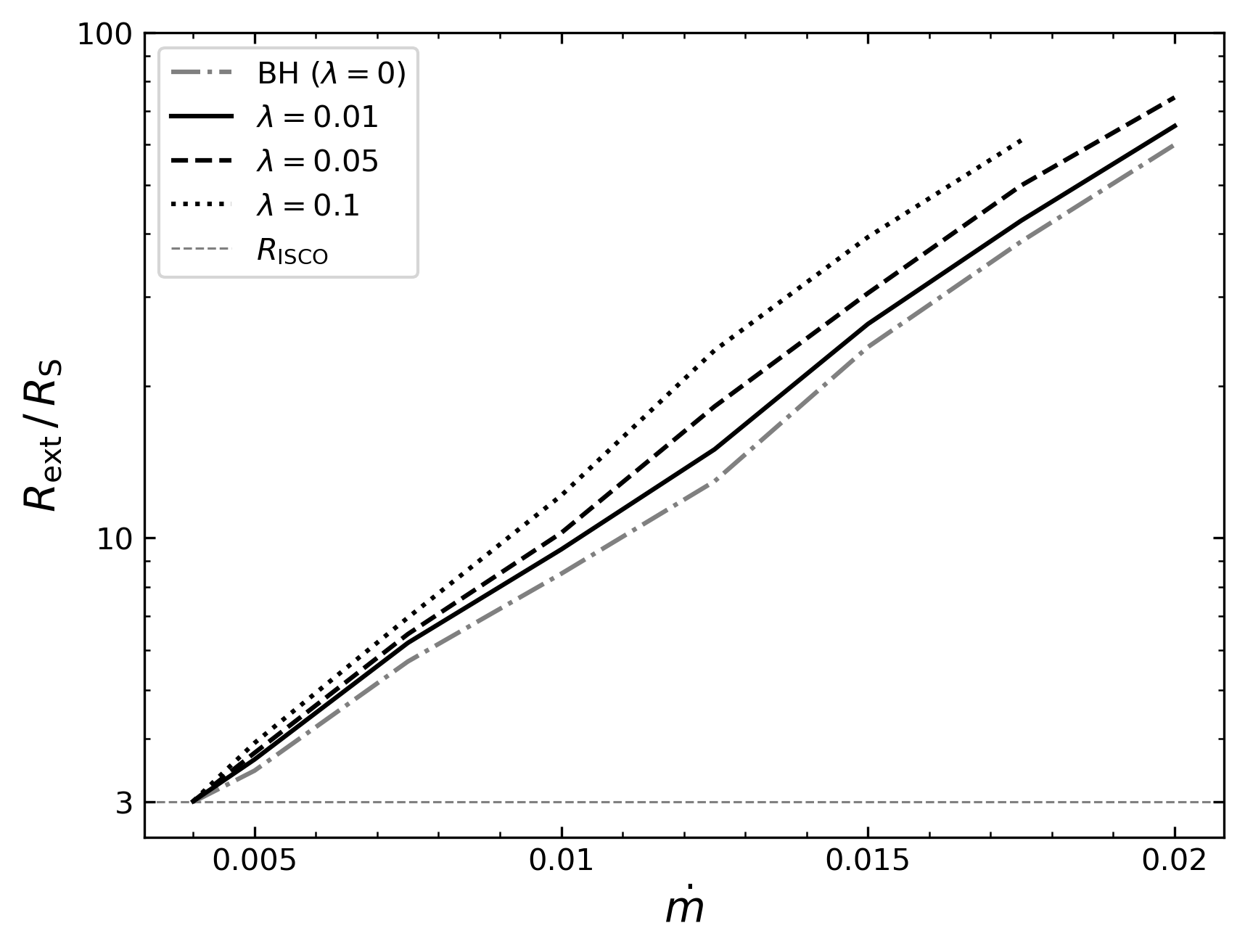}
\caption{Extension of the inner disc as a function of the mass
transfer rate for different values of $\lambda$,  the fraction of
the accretion energy released at the surface of the neutron star to be
thermalised, assumed parameter $\alpha=0.3$. The curve with $\lambda=0$ corresponds to the case of black holes. At very low $\dot m$ the  curves for different $\lambda$ overlap, indicating that the density is too low for Coulomb coupling. }
\label{f:Rextm-lambda}
\end{figure}

With the  chosen parameters we determine the properties of the
accretion flow for a series of mass transfer rates. 
As we are interested in the low hard state, we limit our
  computation for accretion rates up to a few percent of
  the Eddington rate. In this case, the disc
 is transformed to an ADAF by efficient evaporation at
around a few hundred Schwarzschild radii according to the interaction of the disc and corona \citep[e.g.][]{MLM2000L}. When this ADAF  flows
towards the neutron star, re-condensation is expected as a consequence
of radiation cooling in the innermost region, and a weak inner disc
 is formed, fed by continuous condensation. With
increasing accretion rate the condensation becomes more efficient and
 the weak inner disc extends farther outwards.
We show the extension of the disc from the ISCO to the outer
edge in Fig. \ref{f:Rextm-lambda}. As can be seen from the
  figure, the disc size becomes  a few tens of Schwarzschild radii when the total
 accretion rate  $\dot m =\dot M/\dot M_{\rm{Edd}}$ increases to 0.02, and accretion gas via the disc to the
 neutron star increases accordingly. Further increase of the accretion
 rate would finally lead to a contact of the inner and the outer
 disc, and thereby a spectral transition to the soft state.  

The additional soft photons from the neutron star surface make
 possible the existence of an inner disc with
an extension from the ISCO outward to about 70 Schwarzschild
radii for low mass accretion rates. These results clearly show that  in neutron star LMXBs inner discs can exist in 
 an expected hard spectral state due to persistent re-condensation. Our spectral calculations in the following confirm this.

\begin{figure}
\centering
\includegraphics [width=8.5cm]{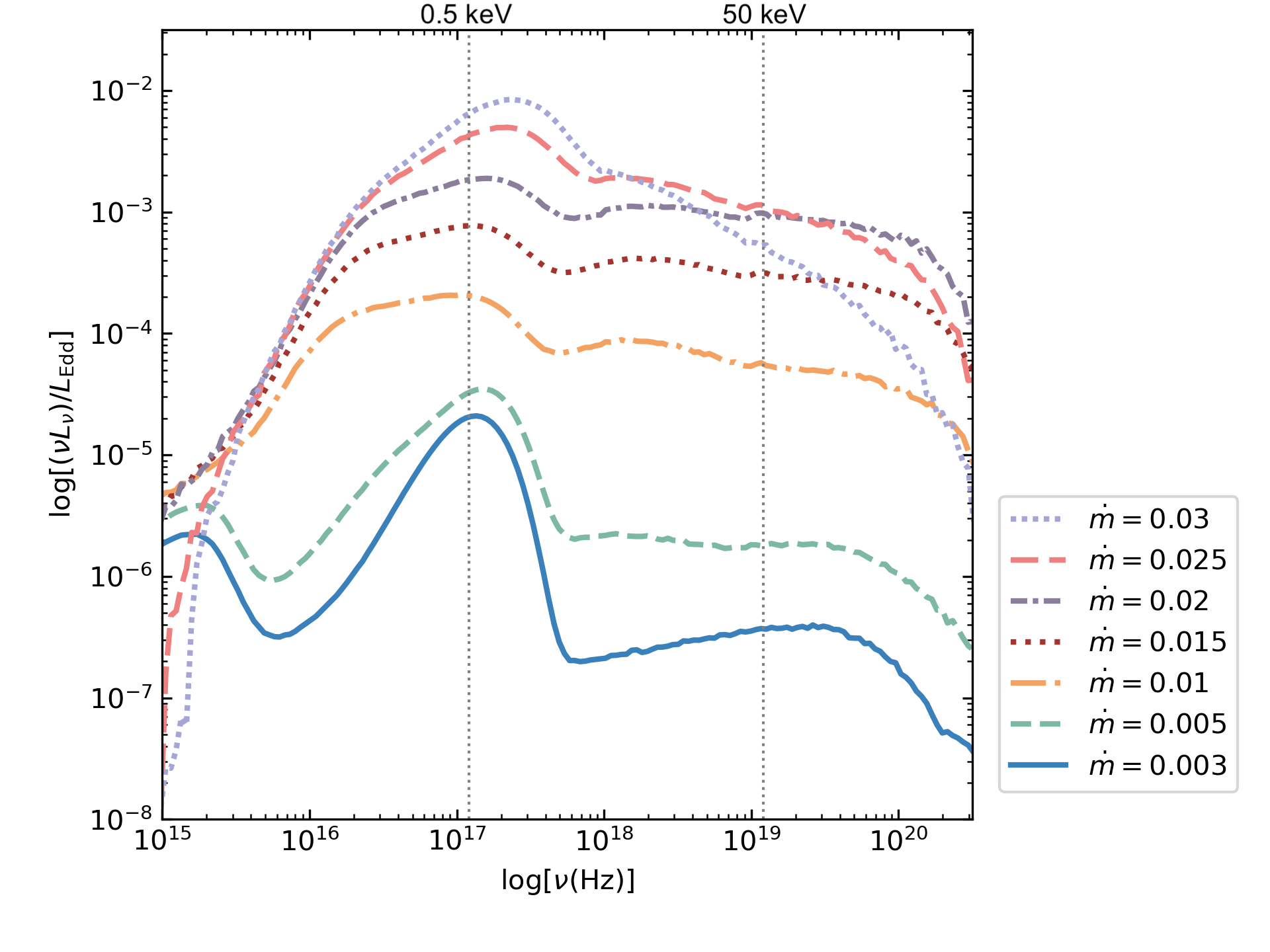}
\caption{Spectra of the two-phase accretion flow
    towards the neutron star for different accretion rates $\dot{m}$,
    assumed parameters $\alpha=0.3$, $\lambda=0.01$. A new component contributed by the neutron star surface is apparently dominant at very low accretion rates, e.g. $\dot{m}=0.003,0.005$, becoming comparable with that from the accretion flows at higher $\dot{m}$, and eventually negligible when the accretion is dominantly via a thin disc.  
    }
\label{f:spectra}
\end{figure}

With the  determined structure of the two-phase accretion flows we
perform Monte Carlo simulations \citep{pozdniakov1977,
manmoto1997, qiao2012,ghosh2013}
to calculate the spectra emergent from the disc and the corona for a
series of accretion rates. Specifically, when the radial distribution of the temperature and density in the corona, the luminosities from the corona and  disc, and the radial distribution of the accretion rate in disc are all determined by the disc-corona interaction model \citep[see][]{wang2024}, we are able to calculate the emissions from  the bremsstrahlung, synchrotron,  and, in particular, the inverse Compton scattering processes by simulations\footnote{The Monte Carlo simulation code, \texttt{HAFspec}, is publicly available at https://github.com/liumingjun-astro/HAFspec. }. As shown in Fig. \ref{f:spectra}, the
overall hard X-ray spectra are similar to a hard state power-law spectrum for
accretion rates $\dot m\lesssim 0.02$. In addition, a  component in
the soft X-ray  band appears, which is the thermal
radiation contributed by the surface/boundary layer of the neutron
star. This component is dominant
at low accretion rates, as shown in Fig. \ref{f:spectra} for $\dot m=0.003, 0.005$. 
At higher
accretion rates, $\dot m=0.01-0.02$, the hard X-ray 
  band is  comparable to neutron star radiation, implying that
  the soft photons from the neutron star surface are no longer predominant (since the corona can
 contribute nearly the same amount of the hard X-rays to soft photons by irradiation and re-procession in the underlying disc). Nevertheless,  it
is still important in the somewhat outer region  (at a few tens of
Schwarzschild radii) where  no disc can exist without  radiation of the neutron star surface. At even higher accretion rates, as shown by the curve for $\dot m=0.03$, the contribution from the neutron star surface is negligible as the disc radiation is much more efficient. 

In brief, the surface radiation does not only change the spectrum by
presenting a new, soft X-ray component and hence enhancing the hard X-ray radiation, but also increases the weak disc extension in the low hard state.
The existence of an inner disc provides the required disc matter for the emission of
the broad iron line. In addition, a transition to the soft state occurs at a slightly lower accretion rate by the  effect of neutron star radiation. It can be seen in Fig. \ref{f:spectra} that the spectrum already corresponds to the soft state 
at $\dot m=0.03$ for the chosen parameter $\alpha=0.3$ and $\lambda=0.01$.

\begin{figure}
\centering
\includegraphics [width=8.5cm]{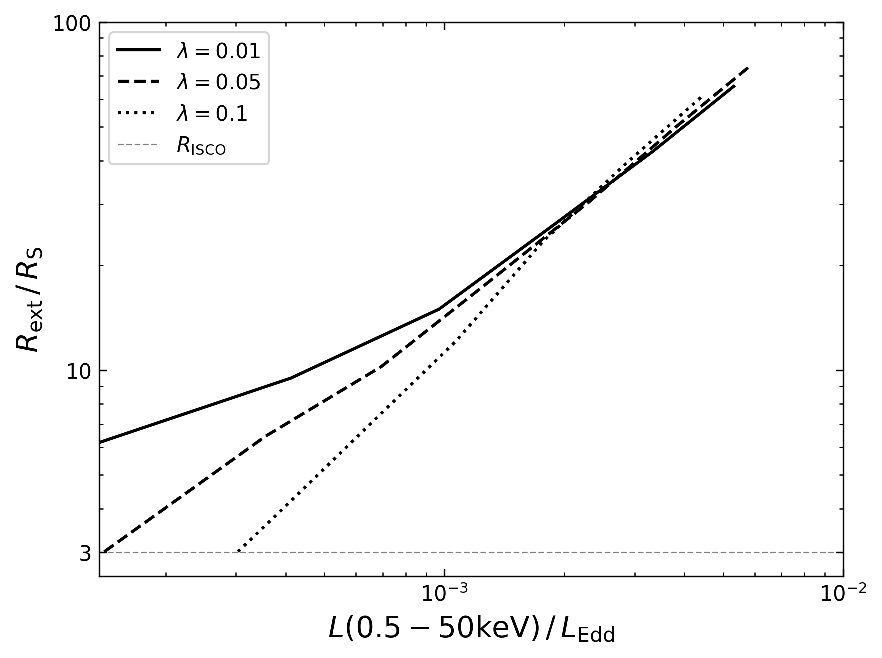}
\caption{Extension of the inner disc as a function of the luminosity $L_{0.5-50{\rm keV}}$ in units of Eddington luminosity, assumed parameter $\alpha=0.3$. A general trend is shown that the disc extension increases with the luminosity. The effect of  $\lambda$ is only obvious  at very low luminosity as it is significantly contributed by the surface of neutron star.
} 
\label{f:Rext-L}
\end{figure}

To compare with observations the extension of the inner disc should be determined as a function of the luminosity.  As  used in the observations for
several sources in the hard spectral state,   the
luminosity in the range 0.5-50 keV is calculated from the theoretical spectra shown in Fig. \ref{f:spectra}. Combined with the results presented in Fig. \ref{f:Rextm-lambda} we  show   in Fig. \ref{f:Rext-L} how 
the extension of inner discs caused by the re-condensation process varies with luminosities. The results indicate that a disc can exist for a
luminosity as low as $L_{\rm 0.5-50 keV}/L_{\rm Edd} \sim 10^{-4}$ depending on the
accretion rate at the hard spectral state. This implies that
relativistically broadened iron lines could appear in a wide  range of
luminosity in the low/hard spectral  state.

  We examine the effect of surface radiation (equivalent to $\lambda$)
on the disc extension.   It is natural that a larger $\lambda$
increases the soft photons and hence the condensation, and leads to a
larger disc for a given accretion rate. However, the effect is not
significant, in particular, at very low accretion rate the coronal
density is so low that Coulomb coupling is too weak and radiation
cooling is anyhow inefficient even with strong soft photons from the
neutron star.  Such results are  demonstrated in
Fig. \ref{f:Rextm-lambda}.  

For a comparison with the black hole LMXBs we set $\lambda=0$ for $m=10$. The computational results, as   plotted in Fig. \ref{f:Rextm-lambda},  indicate smaller disc extension than that in neutron star LMXBs. The difference becomes small at low accretion rates and eventually disappears, further confirming the speculation of too weak Coulomb coupling. Such a result implies that  the lower limit to $\dot m$ for the existence of an inner disc   is similar for both neutron star and black hole LMXBs. Nevertheless, the strength and extension of the inner disc, the luminosity and the spectra in neutron star LMXBs are distinct from black hole LMXBs. 

The effect of $\lambda$ on the relation of disc extension and luminosity (Fig. \ref{f:Rext-L}) is obviously different from the relation of disc extension and accretion rate (Fig. \ref{f:Rextm-lambda}).  This is because an increased  $\lambda$ not only extends the disc but also enhances the luminosity. Such effect is especially stronger at very low accretion rates since the surface radiation contributes a significant fraction to the luminosity (see Fig. \ref{f:spectra}). Therefore, for the same disc
  extension the luminosity is larger for larger $\lambda$. This
  interprets the surprising result as shown in Fig. \ref{f:Rext-L},
  that is, for given  luminosity the disc size is larger if a smaller
  fraction  $\lambda$ of the accretion energy is released to emit in
  the boundary layer of the neutron star.  This relation  only occurs at very low accretion rates  because the radiation from the neutron star no longer contributes a dominant component to the 0.5-50 keV luminosity when the disc and corona become more radiatively efficient at higher accretion  rates.

\begin{figure}
\centering
\includegraphics [width=8.5cm]{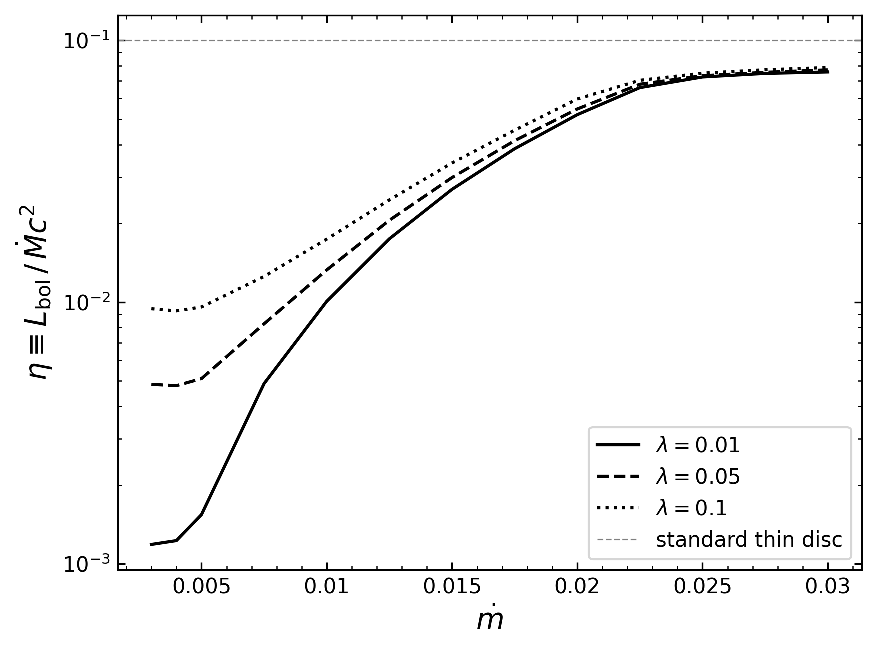}
\caption{Radiative efficiency of the accretion flow to the neutron star,
  including recondensation, as a function of $\dot{m}$, assumed parameter $\alpha=0.3$. The efficiency increases with accretion rate as a consequence of increasing Coulomb coupling; The efficiency is higher for larger $\lambda$ because of more radiation from the neutron star surface and the Compton scattering. }
\label{f:eff}
\end{figure}

 The  radiative efficiency of the accretion flow is  shown in
Fig. \ref{f:eff}, where the emission from the neutron star surface is
also included. A general trend of increasing radiative efficiency with
 increasing accretion rate is demonstrated in the
  figure. This is a consequence of enhanced Coulomb coupling between
  ions and electrons in the corona at high accretion rates. The
  variation at the extremely low accretion rate ($\dot m\lesssim
  0.005$) diminishes with increasing $\dot m$  because the
radiation from the neutron star  exceeds the disc corona radiation,
leaving the efficiency  fairly constant, determined by the surface radiation
 corresponding to $\lambda \times 0.1$ (see Eq. \ref{e:lambda}).  
At high accretion rates, the efficiency shown in Fig. \ref{f:eff} does
not increase  farther until $\eta \sim 0.08$ at $\dot m
\sim 0.03$ where the transition of the spectrum  is expected (see Fig. \ref{f:spectra}). This is consistent with the theoretical value of $1/12$ for a standard disc ending at $R=3R_{\rm S}$. The contribution from the neutron star surface, which is evaluated by $\lambda$ times the luminosity, $\lambda (0.1\dot M c^2)$, is negligible at high accretion rates for  $\lambda=0.01, 0.05,0.1$. For any higher value of $\lambda$,
the  radiation efficiency in a neutron star is no more than double of
the disc, i.e. $1/6$ since the radiation energy is essentially from
the potential energy of the accreted gas, regardless of the detailed
radiation process in the accretion flow \citep[see also][]{qlbf2021}.
The effect of $\lambda$ on the radiation efficiency is distinct at low accretion rates ($\dot m \lesssim 0.02$). The efficiency is higher for larger $\lambda$ because  radiation from both the neutron star surface and the Compton scattering in the corona increases with $\lambda$.

\begin{figure}
\centering
\includegraphics [width=8.0cm]{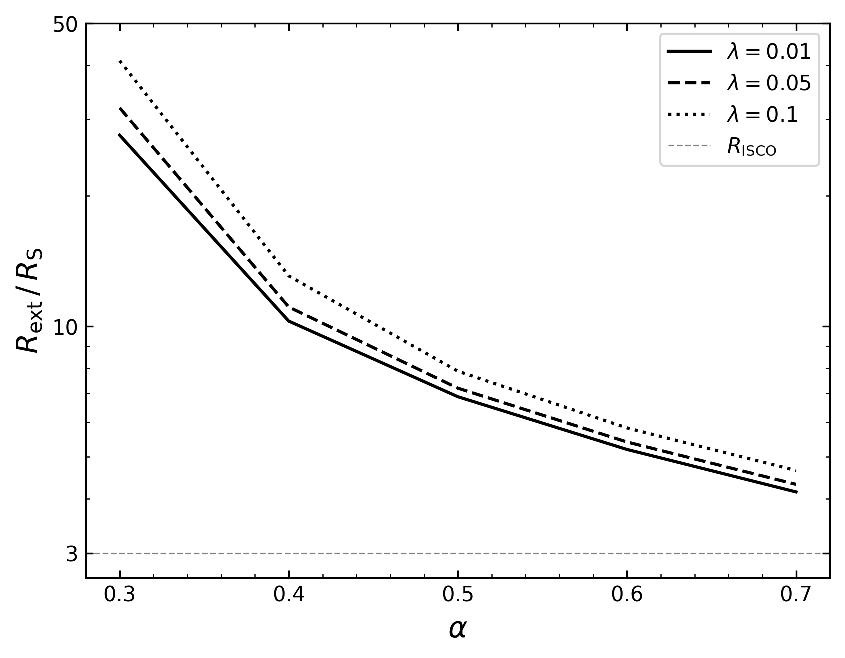}
\caption{Extension of the inner disc as a function of
 the viscosity $\alpha$ for different values of $\lambda$, assumed $\dot m=0.015$. The disc extension decreases with  increasing $\alpha$ as consequences of  the increasing heating and decreasing cooling. }
\label{f:Rextm-alpha}
\end{figure}

Another important role in the accretion flows is the viscosity parameter.
 It affects  many principal properties such as the structure, the instability, the critical accretion rate for the hot flow and the condensation studied here. 
 We show in Fig. \ref{f:Rextm-alpha}  how the disc extension depends on
 the value  of $\alpha$ in the hot flow. As is expected, the discs would be smaller for values of $\alpha$ higher than the preferred value 0.3. Such a feature is similar
  to that of black hole LMXBs. The viscosity parameter affects the
  viscous  heating rate and radial velocity of the accretion flow. An
  increase of viscosity leads to a decrease of density in a steady  (constant $\dot m$) hot
  accretion flow,  thus, the radiative cooling decreases.
Consequently,  condensation is reduced, and  a hot flow can survive at a higher
critical accretion rate. Therefore,  opposite to the
  situation for smaller $\alpha$,  no condensation-fed disc
 exists at a large distance for the same accretion rate.   

\section{Comparison with observations}

\subsection{The existence of inner  discs}

While the classical accretion theory predicts a truncation of the thin
disc  to interpret the hard  spectra, the reflections represented by
relativistically broadened $K_\alpha$ lines observed in low/hard
spectral state call the theory into question.  The
condensation of gas from the inner hot flow allows to sustain a weak disc
close to the central object and thus the occurrence of
broad iron lines.  This condensation
does not change the classical accretion theory, but elucidates
additional physical processes in the accretion flows.  The existence
of the weak inner disc resolves the contradiction between the hot flow
dominated continuum  and  the broad iron lines   as already shown
for black hole LMXBs by \citet[]{wang2024} and is expected to play 
an even more important role in neutron star LMXBs because of the surface radiation. 

X-ray observations of NuSTAR \citep{harr2013}
and NICER \citep{gen2012} provide interesting new information on the
accretion geometry for neutron star LMXBs. 
By reflection modelling the limits
of thin discs in neutron star LMXBs are  independently  determined. 
Derived inner disc radii together with the luminosity 
are listed in several papers \citep{cack2010, eijnden2018, lumb2019, mou2023,
  lud2024}.
In Table \ref{tab:table1} we summarize the derived inner radii $R_{\rm in}$ of the disc available in the literature.
As can be seen from Table \ref{tab:table1}, a common characteristic drawn from these observations is that the reflection components
caused by reprocessed emission from the disc are all from the region
very close to the ISCO  (where $R_{\rm{ISCO}}=3R_{\rm{S}}$), 
even for sources in a very faint hard spectral state. According to the classical theory, the thin disc must be truncated at low accretion rates, $\dot m\lesssim 0.02$.  Even  if a standard thin disc were permitted to extend inward to such small radii, radiation from the accretion flows should be thermal disc-dominated spectra, which is apparently not the case for sources at low/hard state. The existence of a weak inner disc at low accretion rates, as shown in this work,  provides cold matter to emit reflection lines whereas it does not significantly change the continuum in hard X-rays, in good agreement with the observations for the sources at low/hard spectral state.

\subsection{Reflection region  in hard state as compared with the predicted  inner disc} 

 With  the  determination of the reflection region for individual
 sources, the question on the condensation model is  whether the
 predicted  extension of the inner disc  is sufficiently large to cover  the reflection region, 
in particular for sources at very low luminosity. 
The reflection derived inner radii for most
sources are below $2R_{\rm{ISCO}}$ 
  with  Eddington ratios above
0.005 (see Table \ref{tab:table1}).  As can be seen from
Fig. \ref{f:Rext-L}, the  condensation-fed inner discs extend to
more than $R_{\rm ext}>60\,R_{\rm S}$  for $L/L_{\rm Edd}>0.005$. This indicates the condensation-fed inner disc, even if
truncated, can still act as reprocessing media to produce the observed
reflection component.  For 4U 1636-53 with a relatively large
reflection region, the Fe K emission line observed in 2005 was at
about $49R_{\rm g}$ or $24.5R_{\rm S}$ ($R_{\rm g}=GM/c^2$
is the gravitational radius)
with relatively high luminosity, $L/L_{\rm Edd}=0.02$. In the later observation in 2015 a smaller inner disc radius was found at a lower luminosity, $L/L_{\rm Edd}=0.01$.  For the
values of Eddington ratio of 0.01--0.02
an inner disc  is indicated sufficiently large for broad reflection lines 
 as shown in Fig. \ref{f:Rext-L}. 
 
 The faintest source in hard state for which the iron line
could be detected is the very faint X-ray binary (VFXB) IGR
J17062-6143 \citep{eijnden2018}. The two observations for this source document
similar data. The luminosity in 2016 was 16\% lower than that
in 2015 \citep{deg2017} and the inner radius was somewhat smaller than that
derived for the observation in 2015.
  The truncation radius of $77-100R_{\rm g}$ i.e. $38.5-50 R_{\rm {S}}$ with a measured
  Eddington ratio of $0.003$ is  marginally within the condensation
  predicted inner disc. Concerning uncertainties of inner disc radii
  from reflections it should be noted that the values depend on several features, e.g. on the energy band pass of the observations, on the reflection models, on the assumed inclination, distance, mass of the neutron star and the spin parameter \citep{lud2024}.   

Therefore, the observations confirm that the derived reflection regions for
the hard state sources generally lie within the extension of discs
predicted by condensation.

\subsection{Truncation of the inner disc during hard spectral state}

The condensation model provides a physical mechanism for
the existence of an inner disc in the low/hard
state, however, it does not answer the question of how the weak inner disc
is truncated.  As shown in the present study, the interaction
between the hot and cold flows at distances close to the ISCO results in
condensation instead of evaporation. Thus, it is not expected that the
inner disc is truncated by evaporation. 
In neutron star LMXBs a disc truncation can be caused by the magnetic
field of the neutron star. The magnetic pressure controls the matter
flow and thus disrupts the accretion flows at a radius where
the magnetic pressure balances the ram pressure from infalling material
\citep{cack2009}.
This implies that the truncation radius is larger in the case of
stronger magnetic field or smaller accretion rate.  
Transferring the accretion rate into luminosity by adopting an radiation efficiency, the investigation of \citet[][Fig. 9]{eijnden2018}
shows the trend of increasing truncation radius with decreasing
luminosity for a given magnetic field.

The inner radii of discs in  neutron star LMXBs displayed in several diagrams
\citep{eijnden2018, lumb2019, mou2023, lud2024} show no general (one-to-one) trend for all sources in the relation between $R_{\rm in}$
and the luminosity $L/L_{\rm{Edd}}$  \citep[e.g. Fig. 5 of
][]{mou2023}. Such a feature cannot be
 understood if the accretion rate is the sole parameter for
determining the truncation radius. Instead, the truncation by the
magnetic field of the neutron star does not necessarily follow a general relation between the truncation radius and the luminosity because the strength of the magnetic field of a specific source can be different. Therefore, the magnetic truncation of an inner disc is a reasonable scenario in neutron star LMXBs. 

The knowledge of the luminosity and the truncation radius from the reflection studies allows to determine the strength of the magnetic field of the neutron star. Values of the
field strength are listed in Table \ref{tab:table1}, which does show different strength of magnetic field for individual sources.  
 
Magnetic truncation in the hard state is also supported by  the joint {\it NuSTAR} and {\it NICER} observations of 4U 0614 + 091.  
The observations in 2021 and 2022 show a change from soft to hard
spectral state, meanwhile the inner disc receded from ISCO in thermal
dominant spectral state to $1.92R_ {\rm{ISCO}}$ i.e. almost 6$R_{\rm{S}}$ in the power-law
dominant state \citep{mou2023,
mou2024}.
The slight truncation in the hard state could be naturally interpreted by the magnetic field in neutron star LMXBs. An alternative mechanism for small disc truncation 
was suggested as an extended boundary layer on the neutron star \citep{popsun2001}.

The truncation of the inner disc  reduces radiation from the accretion flow. As a consequence, a higher  accretion rate is required in order to 
produce the observational luminosity. This predicts a  larger reflection disc, further supporting 
the interpretation of  broad iron lines originating from the condensation-fed disc.  

\subsection{Disappearance of the relativistically broadened reflection lines}
For  a certain low  accretion rate  gas stops condensation at all distances (see Fig. \ref{f:Rextm-lambda}), predicting that no broad iron lines at low luminosities  (see Fig. \ref{f:Rext-L}).   
In some investigations of neutron star LMXBs it was mentioned that no
Fe K line was found in reflection spectra. \citet{eijnden2020}
described that for the transient source 4U 1608-52 during the rise,
peak, and early/middle decay of the outburst in 2014 a strong, broad
Fe-K$\alpha$ line was seen, but during the 2018 outburst the line
finally disappeared at a luminosity of about 0.002 $L_{\rm{Edd}}$. 
This is in agreement with the model prediction, though a large  viscous parameter ($\alpha>0.3$) is necessary (see Fig. \ref{f:Rextm-alpha}).
Alternatively,  
 the absence of broad iron lines  might  be caused by the strong magnetic fields of the neutron star, which diverts the accretion flows to the magnetic lines at radii beyond the condensation region.  

\begin{table}
 \caption{Sources with inner discs during hard spectral state. Observed radii $R_{\rm{in}}$ are also given in unit of the
     Schwarzschild radius, $R_{\rm{S}}=2GM/c^2$. }
  \label{tab:table1}   
  
  \resizebox{\linewidth}{!}{
  \begin{tabular}{l c l l l l l }    
  \hline                     

  Source&Obs.&$L/L_{\rm{Edd}}$& $R_{\rm{in}}$&$R_{\rm{in}}/R_{\rm{S}}$&$B_{\rm{max}}$&Ref. \\ 
  \hline
IGR J17062-6143  &  2015&  0.003 &   100$R_{\rm g}$  & 50.   &   4.   &     (1)\\
            &  2016&  0.003 &        77$R_{\rm g}$   &38.5   &   2.5  &     (2)\\
	       
4U 1702-429 &  2017&  0.006 &        1.5$R_{\rm {ISCO}}$ &4.5   &    3.3  &     (3)\\
	    
4U 0614+091 &  2013&        &        6$R_{\rm g}$      &3.    &    -        &     (4)\\
            &  2017&   0.007&       1.3$R_{\rm {ISCO}}$   &3.9    &  14.5       &     (3)\\
            &  2021&   0.005&     1.92$R_{\rm {ISCO}}$   &5.7    &   1.2       &     (5)\\

4U 1636-53  &  2005&   0.02 &       49$R_{\rm g}$      &24.5    &    -        &     (6)\\
            &  2015&   0.01 &     1.03$R_{\rm {ISCO}}$   &3.1     &   4.3       &     (7)\\

4U 1543-624 &  2020&   0.013&     1.07$R_{\rm {ISCO}}$   & 3.2     &   0.7       &     (8)\\

1RXS J1804-34  &  2015&   0.02 &  11.1$R_{\rm g}$      &5.6  &   3-1       &     (9)\\

\hline
 \end{tabular}
 }

{\bf{Notes}}. $B_{\rm{max}} (10^8{\rm G})$: magnetic field strength at the pole 
\\References:
(1) \citet{deg2017}, (2) \citet{eijnden2018}, (3) \citet{lumb2019},
(4) \citet{mad2014}, (5) \citet[Obs. 3]{mou2023}, (6) \citet{cack2010},
(7) \citet{lumbb2017}, (8) \citet{lujg2021}, (9) \citet{lumc2016}.

\end{table}

\section{Summary}

The reflection studies in neutron star LMXBs lead to a large number of
derived inner disc radii at quite different luminosities
\citep[Fig. 5]{lud2024}. As generally expected for accretion rates
$\dot{m}\ge 0.02$ (Eddington scaled) a geometrically thin disc reaches inward to
the ISCO and the spectrum is soft, while for low accretion rates ($\dot{m} \le
0.02$) an ADAF fills the inner region and only farther outward a disc
exists, and the spectrum is hard.  The transition is commonly thought to be caused by disc evaporation.
Within this picture the observation of relativistically broadened Fe
lines indicating discs in the innermost region during a hard spectral
state cannot be understood. In this work we elucidate an additional
physical process in the hot accretion flows.  The re-condensation of matter from the coronal flow
to a weak innermost disc allows the produce of relativistically
broadened Fe lines in hard spectral state for neutron star and black hole LMXBs. For black hole binaries the condensation process was
studied by \citet {wang2024}, though less efficient than for neutron
star binaries where the Compton cooling of the corona by the soft
photons of the neutron star enhances the process. 

We determine the extension of the inner discs sustained by
condensation from coronal flows in neutron star LMXBs. For
luminosities $L_{0.5-50{\rm keV}}/L_{\rm{Edd}}$ of 0.0003 to 0.01 the predicted weak
inner discs extend from the center up to 
 almost $100 R_{\rm{S}}$,
   with larger extension for higher Eddington ratio
(see Fig. \ref{f:Rext-L}). 
The disc size is  compared with the observed upper limit of reflection
region in neutron star LMXBs.  With the observational determined
Eddington ratios for specific sources, the theoretically derived inner
discs cover the reflection regions in hard state (see Table \ref{tab:table1}).
These results confirm that the condensation-sustained inner disc serves as the reflection medium at hard state as indicated by broad Fe lines in the reflection spectra.

For systems with extremely low Eddington ratio, the reflection represented by broad Fe lines is predicted to disappear because no condensation occurs to sustain an inner disc at very low accretion rate. In systems with strong magnetic fields,  reflection can also disappear 
since magnetic fields can divert the inner accretion flows  to follow the magnetic lines,  preventing gas from condensation in the inner region. 

Truncation of the inner disc indicated by reflection radius is  thought to be caused by the 
magnetic field of the neutron star. Thus, for  sources with specific strength of magnetic field,  it is not surprising that the inner radii of different sources do not follow a universal relation to the luminosity.

\section*{acknowledgements}
Financial support for this work is provided by the National Natural
Science Foundation of China (Grant No.12333004). This work made use of \href{https://github.com/liumingjun-astro/HAFspec}{\texttt{HAFspec}}, a publicly available Monte Carlo simulation code for the emergent spectrum of hot accretion flows.

\section*{Data availability}
The data and code used in this work will be shared on reasonable
request to the corresponding author.


 \bibliographystyle{mnras}
 \bibliography{nsLMXB} 



\bsp	
\label{lastpage}
\end{document}